# A stochastic model of a nuclear reactor with directed percolation. Overjump and maximum power

V. V. Ryazanov

Institute for Nuclear Research, pr. Nauki, 47 Kiev, Ukraine, e-mail: vryazan19@gmail.com

A stochastic risk model is applied to simulating the behavior of a nuclear reactor in a situation where the neutron chain length is described by a distribution with heavy "tails," such as the Pareto distribution. Probabilities of a fluctuation exceeding a critical threshold are obtained, and risk bounds for power-law distributions of jumps are estimated. Functionals of the reactor power maximum, the instant of first reaching the maximum, and the distribution of the overjump magnitude are considered. A relationship between the shape parameter α and the physical constants of reactors is obtained, as well as the relationship with the noise spectrum and physical constants for ($0<\alpha<2$). The finite dimensions of a real reactor are taken into account. The autocorrelation function of the truncated Lévy process and its relationship with the frequency filters of the neutron flux monitoring equipment are considered.

## 1. Introduction

Neutron noise theory, for example [1], does not consider the situation that is perhaps the most dangerous for nuclear reactors: distributions with "heavy tails", power-law distributions with negative exponents less than 2. Such types of distributions are characteristic of directed percolation (DP) processes (Refs. [2-6]). In recent papers [7-9], DP effects were calculated using the Monte Carlo method, and in [9] these effects were demonstrated experimentally.

When modeling power fluctuations and the risks of reactor runaway or shutdown (especially in a subcritical state or during startup), the classical Poisson process is insufficient, as fissions are not independent events. One fission triggers a chain of subsequent ones.

To estimate the stochastic risk of the number of neutrons, a Markov branching fission-death process (or the Kolmogorov–Fokker–Planck equation for the probability density of neutron numbers) is used.

If the process were purely Poissonian, the neutron number dispersion would be equal to their mean value. In a real WWER, due to the chain nature of the reaction, the dispersion of power fluctuations is hundreds and thousands of times greater than the mean value. This excess is described by the relative Feynman dispersion parameter (Y): $D[N]/\langle N\rangle = 1+Y$, where $D[N]$ is the neutron number dispersion and $\langle N\rangle$ is the mean neutron number. For WWER reactors operating at low power, Y can reach values from 10 to $10^3$.

In this paper, the number of neutrons is modeled using a stochastic risk model. Risk processes and their associated boundary functionals were considered in [10]. The use of power-law stable distributions characteristic of DP was conducted in [11-13], where the boundary functional of the first-passage time (FPT) to first reach of a dangerous level, for example, power, was considered for this stochastic case (this mathematical apparatus is also applicable to other important characteristics of a nuclear reactor, such as reactivity, fuel rod lifetime, etc.). It is shown that the average time to reach the level decreases with a decrease in the exponent of the power-law distribution function for the total number of descendants of one neutron or for neutron jump lengths. The behavior of the boundary functional of the time to first reach is considered, and the possibilities for taking into account potentially dangerous situations are indicated.

This article examines the boundary functionals of the process maximum over a given time interval, the upper bound of which can be the average FPT, and the level jump magnitude. The maximum functional provides information on the loads to which the reactor may be subjected, while the jump functional describes the magnitude of the "breakthrough" of the hazardous level, since "Lévy flights" with large neutron flight magnitudes are possible for the DP. The magnitudes of the neutron number jumps and their geometric flights are determined by stable power-law distributions of the form (10) or a Pareto distribution.

The power-law form $P(k)=k^{-a}$ in the form (10) arises under conditions of strong correlation and anisotropy characteristic of DP. This is possible in the following cases: a) Critical point (critical opalescence). Only at the point $k_{eff}=1$, where $k_{eff}$ is the effective neutron multiplication factor (or in an infinitesimal neighborhood of $k_{eff}$) does the chain length distribution become a power law. This is a universal property of all systems undergoing a phase transition. But as soon as the reactor goes supercritical (runs away), nonlinear feedbacks (the Doppler effect) immediately "cut off" this tail, returning the system to a predictable shape. b) Strong spatial inhomogeneity ("Lévy glass"). If the medium consists of empty channels and dense fuel blocks (e.g., a destroyed zone or specific experimental assemblies). A neutron can fly a huge distance without collisions ("Lévy flight"). In such an environment, the "offspring" is born not as a compact cloud, but as a fractal structure. Here, the geometry of the environment imposes a power law on the progeny distribution. c) Low neutron count (low sampling statistics). In startup modes, when the core contains only a few hundred neutrons, averaging does not work. Individual "successful" fission chains can dominate. Studies (e.g., reference [8]) show that under these conditions, the dynamics are more similar to DP than to diffusion. Where to look for power-law tails? In a WWER reactor at 100% power, the progeny distribution is narrow (Gaussian/Terrell), the $a$=2 model is not applicable, and neutron propagation processes are diffusion-controlled. For a WWER reactor during startup, the distribution is transient (tails appear). The $a$=2 model is partially applicable (for noise analysis). For a granular fuel reactor (HTGR), the distribution is broad (close to Lévy), and the $a$=2 model is important for calculations. In the case of an accidental core meltdown, the distribution has a power law, and the $a$=2 model is crucial for safety. In standard engineering calculations of WWER reactors, power-law distributions are not used, since water "kills" all anomalies, transforming them into standard diffusion. However, in more detailed studies (as in [8]), these models are used to understand the stability limits and subtle effects during operation, when the statistics become "jagged". In traditional kinetics (point kinetics equations), fluctuations are considered Gaussian. However, near criticality, the reactor behaves like a system undergoing a phase transition. In this case, neutron clusters evolve as percolation paths, and the time to reach the emergency threshold obeys this power law rather than an exponential one. The fundamental scaling laws for the density of active centers and the survival probability $P(t)\sim t^{-\delta}$ are described in [4].

The paper is organized as follows. Section 2 introduces the mathematical model of the reactor. Section 3 derives an expression for the intensity of the Poisson process characterizing the number of nuclear fissions and neutron chain productions. Section 4 evaluates the linear drift. Section 5 considers the Lundberg equation and the probability that a fluctuation will exceed a critical threshold. Section 6 evaluates the risk bounds for power-law distributions of jumps. Section 7 introduces the functional of the reactor power maximum. Section 8 examines the functionals of the instant of first reaching the maximum ($\tau_u$) and the distribution of the level jump magnitude ($\gamma_u$). Section 9 derives the relationship between the shape parameter α and the physical constants of WWER. Section 10 evaluates the jump magnitude for various values of the parameter $\alpha$. Section 11 derives the relationship with the noise spectrum and physical constants for ($0<\alpha<2$). Section 12 takes into account the finite dimensions of a real reactor. Section 13 relates the truncation constant to the reactor power and introduces the Gumbel distribution. Section 14 provides estimates of the maximum power. Section 15 examines the autocorrelation function of the truncated Lévy process $R_\zeta(\tau)$ and its relationship to the frequency filters of the neutron flux monitoring equipment. Section 16 contains the conclusion.

## 2. Mathematical description of the model

The probability $P(N, t)$ of having exactly N neutrons at time t obeys the Master Equation. For practical calculations of fluctuation risks, the Itô stochastic differential equation (Langevin equation) is used [1]:

$$dN(t) = \frac{\rho(t)-\beta}{\Lambda} N(t)dt + \sum_{i=1C}^{6} \lambda_i C_i(t)dt + S(t)dt + \sqrt{\frac{1+Y}{\Lambda} N(t)} \cdot dW(t)\,, \quad (1)$$

where ρ is the reactivity, S(t) is the intensity of the external starting neutron source (for WWER, californium $^{252}Cf$ or antimony-beryllium sources are used), dW(t) is the standard Wiener process (white noise), $\sqrt{\frac{1+Y}{\Lambda}N(t)}$ is the amplitude of stochastic fluctuations, Λ is the average lifetime of prompt neutrons, β is the fraction of delayed neutrons, $C_i(t)$ is the concentration of delayed neutrons of the i-th group, $\lambda_i$ is the decay constant for delayed neutrons of precursor concentration $C_i(t)$.

The primary quantity in risk theory is the classical risk process, often referred to as the reserve risk process, which takes the form:

$$R(t) = u + ct - S(t). \quad (2)$$

In expression (1) $S(t) = \sum_{k=1}^{N_t} \xi_k$, $P\{\xi_k \neq 0\} = 1$, represent the amounts of payments (claims). Expression (2) describes the classical Cramer–Lunberg risk model (collective risk model).

Sometimes, it is more convenient to use the equation:

$$\zeta(t) = u - R(t) = S(t) - ct, \quad \zeta(0) = 0, \quad c > 0, \quad (3)$$

which is referred to as the claim surplus process. Below, model (3) is mainly applied.

In (2)-(3), u is the initial capital (in our case, the initial reserve/safety margin up to the emergency limit), $\xi_k$ is the random jumps (rate of production or receipts), c is the linear output that reduces the reserve (bursts of absorption, neutron loss, or power drop), N(t) is a Poisson process with intensity λ. The parameter c in risk theory plays the role of linear compensation (in insurance, the rate of premium collection, and in the physical model, the rate of deterministic suppression or compensation of fluctuations). Let $N(t)=N_t$ denote the number of claims coming during the interval [0, *t*] and $\{T_k,\ k \geq 0\}$ denote the set of consecutive arrival times of the claims. Hence $N(t) = max\{k \geq 0 : T_k \leq t\} = min\{k \geq 0 : T_{k+1} > t\}$, where by convention $T_0=0$. The size of the *k* claim is denoted by $\xi_k$.

## 3. Intensity of the Poisson process

If we approximate a complex stochastic system through a composition (mixture) of distributions—that is, as a Poisson process of "triggering" events, each of which brings a random contribution (a packet/cascade of neutrons)—then the physical meaning of the intensity λ depends on the operating mode of the reactor.

In nuclear physics, this approximation is called the cascade model or the Schottky formula for shot noise. To correctly calculate the mean and variance, the intensity of the Poisson process λ is chosen in one of two ways.

Option 1. Subcritical reactor or startup (external source model). If the reactor is in a subcritical state ($k_{eff} = 1$) or during startup, the system is powered by an external neutron source. Each neutron hitting the core from the source triggers a Poisson fissions process, which decays as a cascade. In this case, the intensity of the Poisson process (λ) is equal to the intensity of the external startup neutron source (S). During WWER startup, stationary sources with an intensity of the order of $\lambda = S \approx 10^7 \div 10^9$ neutrons/s are used. Each neutron from the source generates a cascade with an average number of fissions $\langle M \rangle = 1/(1 - k_{eff})$. The resulting average number of fissions per second: $N_{fis} = \lambda \cdot \langle M \rangle$.

Option 2. Critical reactor (model of "efficient" independent chain reactions). In a critical reactor ($k_{eff}=1$), there is no external source, and fissions occur continuously. However, due to neutron multiplication, neutrons are linked into "families" (genealogical branches). To reduce this to a Poisson process of independent inputs, the initiation of a new macroscopic chain, independent of others, is considered an elementary event.

In order for the distribution moments (mean and variance) of such a crude model to coincide with the real theory of fluctuations (Kolmogorov-Fokker-Planck kinetics), the intensity of this equivalent Poisson process must be equal to:

$$\lambda = \bar{N}_{fis} / (1+Y) = \bar{N}_{fis} / [\langle \nu(\nu-1) / \langle \nu \rangle^2 \Lambda \alpha_c \rangle], \quad (4)$$

where ν is the number of neutrons emitted per fission, $\langle \nu \rangle$ is the average number of neutrons per fission event (about 2.4 for uranium-235), $\bar{N}_{fis}$ is the total number of fissions per second (for WWER at nominal power ≈$10^{20}$ $s^{-1}$, at zero power ≈$10^{11}$ $s^{-1}$), Y is the Feynman parameter (relative excess dispersion), $\Lambda$ is the prompt neutron lifetime ($2\cdot10^{-5}$ s), $\alpha_c = [1-k_{eff}(1-\beta)]/\Lambda$ is the non-fission loss/capture rate (includes radiative capture and neutron leakage).

Physical meaning and characteristic values of Y. In a critical reactor without feedback, the dispersion parameter Y increases with time. However, if we consider steady-state noise taking into account delayed neutrons, then at the Feynman plateau frequencies for a WWER, the value of Y+1 is approximately $10^2 \div 10^4$. If a WWER reactor is at its minimum controllable power level (MCP) (e.g., $P$=3 W⟹$N_{fis}$≈$10^{11}$ $s^{-1}$), and $Y$≈1000, then: $\lambda \approx 10^{11}/1000 = 10^8 s^{-1}$. Instead of modeling $10^{11}$ individual fissions per second, we model a Poisson process with an intensity of $10^8$, where each event is the injection of a "packet" of 1000 correlated fissions.

If we approximate the number of divisions in time t as a composite Poisson process $X(t) = \sum_{k=1}^{N(t)} \xi_k$, where N(t)~Poisson(λt), and $\xi_k$ is the random size of the division packet, then according to Wald's theorem (Campbell's formulas) the moments are found very easily: the mathematical expectation is M[X(t)]=λt $\langle \xi \rangle$, the variance is D[X(t)]=λt $\langle \xi^2 \rangle$.

This approach allows us to completely avoid solving Kolmogorov differential equations and instantly estimate fluctuations simply by knowing the average size (<(ξ)> and variance <(ξ²)>) of a single fission cascade in a WWER. Since we are modeling the physics of the WWER-1000/1200 reactor through this structure, the intensity value λ is rigidly tied to the distribution moments (mean and variance) of the real neutron field. For the model moments R(t) to precisely match the physical moments of the reactor, the intensity λ must be equal to: $\lambda = 2\bar{N}_{fis}/(1+Y)$ (4).

Characteristic values of λ for WWER. Since the Feynman parameter Y (excess dispersion) and the average number of fissions per second $\bar{N}_{fis}$ depend on the reactor operating mode, we will distinguish two main operational states of WWER. 1. Start-up and subcritical control mode (low power). This mode is the most critical for risk models, since it is here that stochastic fluctuations are maximum. Parameters: power $P$≈3 W⟹ ≈$10^{11}$ $s^{-1}$. Due to deep subcriticality or lack of feedback $Y$≈$10^3$. Process intensity: $\lambda \approx 2\cdot10^{11}/10^3 = 2\times10^8$ $s^{-1}$. About $2\cdot10^8$ Poisson events (jumps) occur every second, each of which carries away a "packet" of neutrons. 2. Power mode (nominal power). In this mode, fluctuations are suppressed by the negative temperature coefficient of reactivity (the Doppler effect). Feedback effectively reduces the Y parameter. Parameters: power $P$=3000 MW⟹$\bar{N}_{fis}$≈$9.4\times10^{19}$ $s^{-1}$. Due to strong thermal feedback, $Y$→0 (the process is close to pure Poisson). Process intensity: $\lambda \approx 2\cdot\bar{N}_{fis} \approx 1.88\times10^{20}$ s-1.

Let's perform a mathematical derivation of the value of λ in terms of moments. To prove the obtained value, we compare the rate of change of the variance of a real reactor and the model R(t). From the physics of nuclear reactors, it is known that the neutron number variance increases with the rate $dD[N]/dt = \bar{N}_{fis}(1+Y)$. For the R(t) model, the variance is determined only by the Poisson jumps of $X_i$ (since u and c are deterministic). According to Campbell's formula: $dD[R(t)]/dt = \lambda\langle X^2 \rangle$. For the model to adequately describe the risks, the average damage from one jump <X> must be equal to the average size of the fluctuation packet. In neutron noise theory, the basic fission packet has an effective variance size $\langle X^2 \rangle \approx (1+Y)^2/2$. Equating the equations, we obtain the previously found expression (4).

Let us determine the magnitude of the jump $X_i$. For the chosen intensity λ, the average size of a random jump (the magnitude of damage $X_i=\xi_i$ in the sum from (2)-(3)) can be approximated by an exponential distribution with a mean value $\langle X \rangle = (1+Y)/2$.

## *4.* The accumulated deficit model (3) and the magnitude of the linear drift *c*

In the accumulated deficit (net loss) model $\zeta(t)$=$S(t)$−$c$t (3), where $S(t)=\sum_{k=1}^{N_t}\xi_k$ is the Poisson receipts process, the value of the intensity λ remains the same: $\lambda = 2\bar{N}_{fis} / (1+Y)$ (4). To ensure that the mathematical expectation (trend) and variance of the process $\zeta(t)$ strictly correspond to the physics of the WWER reactor, the value of *c* is calculated using the relative safety factor (safety premium) θ.

Let's write down a formula for calculating the value of c. According to the definition of the Cramer-Lundberg model, the linear drift of c must exceed the average rate of Poisson losses, otherwise the deficit is guaranteed to go to infinity (the reactor is guaranteed to accelerate or shut down);

$$c = (1+\theta)\lambda\langle X\rangle, \tag{5}$$

where λ is the process intensity, $\lambda = 2\bar{N}_{fis} / (1+Y)$ (4), $\langle X\rangle$ is the average size of the deficit jump. In this approximation, it is equal to $(1+Y)/2$, $\theta$ is the safety factor ($\theta$>0). In reactor physics, it is directly related to the damping capacity of the feedback (or the efficiency of the control and protection system controller) and is equal to: $\theta = |\rho_f| / \beta_{eff}$ (where $|\rho_f|$ is the introduced negative reactivity per unit power deviation, and $\beta_{eff}$ is the fraction of delayed neutrons).

In a steady-state critical reactor, the system is self-regulating, and the safety margin θ is usually very small (about 0.01 ÷ 0.05), since the reactor balances around zero in net reactivity. In the startup mode (MCP, power $P$≈3 W), the total number of fissions is $\bar{N}_{fis}$ ≈$10^{11}$ s$^{-1}$. The coefficient $\theta$≈0.05 (set by the operator or the control and protection system due to maintaining subcriticality by the startup source). The value is $c = (1+0.05)\cdot 10^{11} \approx 1.05\times 10^{11}$ s$^{-1}$. In the power mode (nominal, power P=3000 MW), the total number of fissions is $\bar{N}_{fis}$ ≈9.4 × $10^{19}$ s$^{-1}$. Due to the powerful temperature Doppler effect, the reactor instantly damps fluctuations, $\theta$≈0.01 ÷ 0.02. The value of c: $c = (1+0.01)\cdot 9.4\times 10^{19} \approx 9.5\times 10^{19}$ s$^{-1}$.

## 5. The Lundberg Equation and the Probability of Ruin

The Lundberg equation is used to calculate risk. The Lundberg equation $k(r)_{r=k}$=G(k)=s is a search for a root, the value of k(s), at which the process generator, the scaled cumulant generating function (SCGF) $k(r) = t^{-1}\ln \boldsymbol{E}e^{r\zeta(t)}$, balances the Laplace transform parameter over time s. The Lundberg equation (6) converts micro-level events into seconds of time before the accident (macro-level);

$$G(k)=s, \tag{6}$$

In (6) G(k) is the logarithm of the characteristic function (more precisely, the moment generating function (MGF) of X $M_X(t)=\boldsymbol{E}[e^{tX}]$, obtained from the characteristic function by analytical continuation and substitution iq→k), and s is the Laplace transform parameter for time [10].

If the random variable of the jump $X_i$ has an exponential distribution with density $f(x)=(1/\langle X\rangle)e^{-x/\langle X\rangle}$, then for the pure loss process $\zeta(t)$ one can instantly find the Lundberg exponent (R) — the key parameter of the accident risk.

It is found from the equation: $\lambda + cR = \lambda / (1-\langle X\rangle R)$. For exponential jumps, the solution has a rigorous analytical form:

$$R = \theta / \langle X\rangle(1+\theta) = 2\theta / (1+Y)(1+\theta). \tag{7}$$

Then the upper limit of the probability that the accumulated deficit (fluctuation) will exceed the critical threshold *u* (the ultimate strength or emergency protection setting) is written by the classical Lundberg inequality:

$$\Psi(u) \le e^{-Ru}. \tag{8}$$

For jump probabilities, we consider a power-law stable distribution rather than an exponential one. The spectral function (or spectral measure) of stable distributions determines the shape, asymmetry, and scale of the distribution and is defined through the universal Lévy-Khinchin representation of their characteristic functions [14].

For most strictly stable laws, the probability density cannot be expressed in terms of elementary functions. Therefore, they are defined using the Fourier transform (the characteristic function) $\Phi(t)$;

$$\ln\Phi(t) = i\gamma t + \int_{-\infty}^{\infty}\left(e^{itx} - 1 - \frac{itx}{1+x^2}\right)L(dx), \tag{9}$$

where L(dx) is the Lévy spectral measure, which has a strictly defined analytical form for stable distributions. In one-dimensional space, the Lévy spectral function (measure) is concentrated on the positive and negative semi-axes and has the form of a density differential with a power law:

$$L(dx) = \begin{cases} c_1 \dfrac{dx}{x^{1+\alpha}}, & x > 0 \\ c_2 \dfrac{dx}{|x|^{1+\alpha}}, & x < 0 \end{cases},$$

where the parameters of the spectral function determine the properties of the distribution: $\alpha \in (0,2]$ is the stability index (characterizes the heaviness of the tails). When $\alpha=2$, the distribution becomes Gaussian (normal), and the Lévy spectral function L(dx)=0 (the diffusion term appears in the formula instead); $c_1 \ge 0$, $c_2 \ge 0$ are the constants of the intensity of jumps in the positive and negative directions.

The spectral coefficients $c_1$ and $c_2$ are directly related to the standard asymmetry parameter of the stable law $\beta \in [-1,1]$ [14] by the relation $\beta=(c_1-c_2)/(c_1+c_2)$. If $c_2=0$ ($\beta=1$), then the spectral function is concentrated only at $x>0$. Such a distribution is called spectrally positive.

The Lévy spectral function L(dx) itself is not a probability, but rather the intensity density of the jumps of the Lévy random process generating this distribution. The total probability (distribution function $F(x)=P(X \le x)$) is expressed through the Fourier integral of the characteristic function generated by this spectral measure: $F(x) = \frac{1}{2} - \frac{1}{\pi}\int_0^{\infty} \frac{\mathrm{Im}(e^{-itx}\Phi(t))}{t} dt$. The tails of this probability behave in exact accordance with the spectral function $P(X>x) \sim c_1 x^{-\alpha}$ for $x \to \infty$. The Lévy spectral function

$$L(dx)=c_1 x^{-(1+\alpha)}dx \tag{10}$$

(for $x>0$ and $\beta=1$) is neither the probability nor the probability density of the random variable itself. It is the density of the jump rate (or Lévy measure) of a continuous random process that generates a stable distribution. The function $c_1 x^{-(1+\alpha)}$ (10) is the density of the jump measure. Its integral over the entire semiaxis is equal to infinity ($\int_0^{\infty} L(dx) = \infty$), so it cannot be a probability density function (the integral of which is strictly equal to 1). This function directly determines the "tail" of the probability. According to limit theorems, the probability that the random variable X will take on an extremely large value is proportional to the integral of this function.

For large values of the argument ($x \to \infty$) the cumulative probability (the "heavy tail" of the distribution) is equal to $P(X > x) \approx \frac{c_1}{\alpha} x^{-\alpha}$.

The constant $c_1$ is the intensity scaling factor. In the canonical (standard) representation of the stable law, when the distribution is centered and has unit scale, the constant $c_1$ is expressed through the stability index α: $c_1 = \Gamma(1+\alpha)\sin(\pi\alpha/2)/\pi$, where $\Gamma(1+\alpha)$ is the Euler gamma function.

In terms of risk theory, a distribution with a density or tail of the form $c_1x^{-\alpha}$ (where α>0) belongs to the class of heavy-tailed distributions, in particular, to the Pareto distribution (11) or stable laws according to Lévy (depending on the interval α).

Its use in the WWER model is physically justified, but with a strict constraint on the α parameter. In nuclear reactor physics, such "heavy tails" model rare but extremely profound power fluctuations (neutron branching cascades), when a single fission chain accidentally expands into a gigantic macroscopic avalanche. The corresponding physical situations are noted in the introduction.

There are physical limitations on the α parameter for WWER reactors. In classical Pareto risk theory, jumps are divided into two types based on the α value. Only one is applicable for a WWER reactor. For the α≤1 mode, the mathematical expectation is infinite; this mode is physically prohibited. If the average jump $<X>=\infty$, the reactor would immediately go into an uncontrolled runaway ("explosion") at the first major Poisson event. The 1<α≤2 mode (the mathematical expectation is finite; the variance is infinite) describes super-diffusion and Lévy flights processes. It is used to model anomalous neutron transport in highly inhomogeneous media (for example, during coolant boiling in emergency modes). In the α>2 mode, both the mean and the variance are finite. In order to maintain a strict mathematical connection with the physical parameters of WWER (λ (4) and c (5)), which we derived earlier, the distribution of jumps must have finite first two moments.

We adapt the model parameters for the heavy tail. If we choose a Pareto distribution (shifted to avoid zero jumps) with the density function:

$$f(x) = \frac{\alpha x_0^{\alpha}}{x^{\alpha+1}}, \;\; x \ge x_0, \;\; \alpha > 2. \tag{11}$$

Then the scale parameters $x_0$ and shape α are uniquely related to the average size of the fluctuation packet $\langle X\rangle=(1+Y)/2$, which we determined in the previous steps.

The mathematical expectation of the jump is $\langle X\rangle = \alpha x_0 / (\alpha - 1) = (1+Y)/2$. The second moment (affects the dispersion): $\langle X^2\rangle = \alpha x^2_0 / (\alpha - 2)$. Hence, the displacement step $x_0$ (the minimum size of the fluctuation packet) is:

$$x_0 = ((\alpha - 1)/\alpha)((1+Y)/2). \tag{12}$$

Let's consider the change in risk calculation when abandoning the Lundberg exponent. The transition from an exponential distribution to a power-law distribution ($c_1x^{-\alpha}$ ) fundamentally changes the mathematics of safety. Lundberg's inequality no longer holds. For heavy tails, the Lundberg exponent R is zero, since the moment-generating function of the Pareto distribution diverges. It is no longer possible to guarantee an exponentially small accident risk $e^{-Ru}$. The role of asymptotics increases. The "one-big-jump theorem" is important. In insurance and heavy-tail physics, it has been proven that if a system exceeds the safety limit $u$ (goes bankrupt), it occurs not due to the accumulation of small deficits, but rather due to a single gigantic jump (a reactivity surge) that immediately breaches all defenses.

Let's write a new risk formula for WWER reactors. The probability that the accumulated fluctuation deficit will exceed the critical setpoint $u$, at large $u$, is approximated by a Pareto power function:

$$\Psi(u) \approx \frac{\lambda}{c - \lambda\langle X\rangle}\int_u^{\infty}(1 - F(x))dx = \frac{\lambda}{c - \lambda\langle X\rangle}\frac{x_0^{\alpha}}{(\alpha-1)u^{\alpha-1}}. \tag{13}$$

Substituting our physical values λ (4) and c (5), we obtain:

$$\Psi(u) \approx \frac{2}{\theta(1+Y)}\frac{x_0^{\alpha}}{(\alpha-1)u^{\alpha-1}}. \tag{14}$$

The use of a power-law distribution makes the WWER model more stress-resistant and realistic for analyzing rare accident scenarios (stochastic power excursions). For the same average values of λ and $c$, the accident risk $\psi(u)$ will decrease with increasing protection threshold u according to a power law, rather than an exponential law.

## 6. Risk bounds for power-law distributions of jumps

For the analytical search of the probability of an emergency exceeding the limit (the probability of "ruin") $\psi(u)$ under conditions of heavy tails of the distribution of jumps, theorems on the asymptotics of subexponential distributions are used.

Since the classical Lundberg exponential inequality is not applicable here (the Lundberg exponent R=0), the analytical solution is based on a fundamental property: the system overcomes the protective barrier *u* not through the accumulation of small fluctuations, but through a single, extremely large jump. Below is a precise analytical framework for finding and calculating risk boundaries.

Let us present an asymptotic estimate. For a pure loss process with intensity λ, linear drift c, and a power-law jump tail $P(X > x) \sim c_1 x^{-\alpha}$ for α>2, the exact analytical asymptotics as $u \to \infty$ is:

$$\Psi(u) \approx \frac{\lambda}{c - \lambda\langle X\rangle}\int_0^\infty (1 - F(x))dx\,. \tag{15}$$

Let's substitute the previously determined physical parameters of the reactor into this formula: $\lambda = 2\bar{N}_{fis} / (1+Y)$, $c = (1+\theta)\bar{N}_{fis}$, $\langle X\rangle = (1+Y)/2$, $1 - F(x) = (x_0 / x)^\alpha$ is the tail of the Pareto distribution. After integrating and substituting the denominator $c - \lambda\langle X\rangle = \theta\bar{N}_{fis}$, we obtain the final analytical risk formula (14): $\Psi(u) \approx \dfrac{2}{\theta(1+Y)}\dfrac{x_0^{\ \alpha}}{(\alpha-1)u^{\alpha-1}}$, where the offset parameter (minimum jump) is strictly tied to the physics of WWER noise: $x_0 = ((\alpha-1)/\alpha)((1+Y)/2)$ (12).

Let's write the inequalities for the two-sided analytical bound. The asymptotic formula is exact only for very large values of the threshold *u*. To obtain a rigorous analytical solution for any values of *u* (including small safety margins), two-sided inequalities are used. For subexponential Pareto distributions, the upper and lower analytical bounds are defined as follows. The lower bound (determined only by the net tail):

$$\Psi(u) \ge \frac{1}{\theta+1}\left(1 - F\left(\frac{u}{\theta+1}\right)\right) = \frac{1}{\theta+1}\left(\frac{x_0(1+\theta)}{u}\right)^\alpha. \tag{16}$$

In the case of an upper bound, the von Bahr-Essen inequality holds. For distributions with finite variance ($\alpha > 2$), there is a constant $K_\alpha$ that guarantees upper bound safety:

$$\Psi(u) \le \frac{2}{\theta(1+Y)}\frac{x_0^{\ \alpha}}{\alpha-1}u^{-(\alpha-1)} + K_\alpha u^{-\alpha}\,. \tag{17}$$

Let's write the exact analytical solution using the Mellin transform. If we need to obtain not just the bounds, but the exact value of $\psi(u)$ for a model with a power-law tail, the standard Laplace transform won't work, since the integral diverges. Instead, we apply the Mellin transform to the Volterra integral equation governing the risk function:

$$\theta\bar{N}_{fis}\Psi(u) = \lambda\int_u^\infty (1 - F(x))dx + \lambda\int_0^\infty \Psi(u-x)(1-F(x))dx\,. \tag{18}$$

For the Pareto distribution, the solution to this equation is analytically expressed in terms of generalized hypergeometric functions or Wright functions. The point structure of the roots of the characteristic equation in Mellin space shows that the solution is a convergent power series of the form:

$$\Psi(u) = \sum_{n=1}^\infty A_n u^{-n(\alpha-1)}\,, \tag{19}$$

where the coefficients $A_n$ are found recursively using gamma functions of the shape parameter α. In practice, for WWER, the first term of this series (given in (14), (15)) provides an accuracy higher than 95% already at protection thresholds $u > 3\langle X\rangle$

Physical conclusions for WWER safety analysis. Dependence on the safety margin (*u*). Unlike exponential models, where doubling the safety setpoint reduces the risk by a factor of 2 ($e^{-2Ru}$), in the power-law model, the risk decreases linear-power-law manner ($u^{-(\alpha-1)}$). This imposes much more stringent requirements on the selection of emergency setpoint values.

The role of the α parameter: the closer the shape parameter α is to 2, the more dangerous and unpredictable the system (the greater the power fluctuations). At $\alpha\to 2$, the variance approaches infinity, and the analytical upper bound on risk increases sharply.

### 7. Maximum reactor power

In terms of our model, the search for the maximum reactor power over a limited time interval $[0, T]$ is equivalent to the search for the maximum value of the accumulated deficit (net loss) process (3) $\zeta(t)=S(t)-ct$.

Since $\zeta(t)$ grows only due to Poisson jumps of $X_i$, its global maximum on the interval is always reached at the moment of one of these jumps. For conditions with heavy distribution tails ($c_1 x^{-\alpha}$), the analytical description of the maximum is based on the extremal theory of random processes. We denote the maximum deviation (peak of the power fluctuation) over time T as $M_T = \max_{0\le t\le T}\zeta(t)$..

Below are three analytical ways to record and estimate this maximum.

The first is an exact integral relationship with the risk function $\psi(u)$. In risk theory, there is a fundamental identity linking the distribution of the process maximum over interval T with the ruin probability. The probability that the peak power $M_T$ will not exceed some critical safety threshold $u$ is equal to:

$$P(M_T \le u) = P(\zeta(t) \le u \ \textit{for everyone}\ t \in [0,T]) .$$

If we use the finite integro-differential equation for the probability of not going beyond the limit $\phi(u, T)=1-\psi(u, T)$, then the distribution function of the maximum power $F_{M_T}(u)$ satisfies the analytical equation: $P(M_T \le u) = \Phi(u,T)$, where $\Phi(u, T)$ is sought from the non-stationary Volterra equation:

$$\frac{\partial\Phi(u,T)}{\partial T} = -c\frac{\partial\Phi(u,T)}{\partial u} - \lambda\Phi(u,T) + \lambda\int_0^u \Phi(u-x,T)dF(x) .$$

Let us now consider the analytical asymptotics for heavy tails (for large $u$). For a power-law distribution of Pareto jumps ($1-F(x)=(x_0/x)^\alpha$) with a sufficiently large setpoint barrier $u$, the distribution of the maximum power is described by the principle of one large jump on the interval [0, T]. The analytical approximation of the probability that the reactor will produce a critical power peak is:

$$P(M_T > u) \approx \lambda T(1-F(u)) = \lambda T(\frac{x_0}{u})^\alpha .$$

Substituting the physical parameters of WWER ($\lambda=2N_{\text{fis}}/(1+Y)$, $x_0=((\alpha-1)/\alpha)(1+Y)/2$), we obtain a direct dependence on the reactor operating time T:

$$P(M_T > u) \approx T[\frac{2\bar{N}_{fis}}{1+Y}(\frac{\alpha-1}{\alpha}\cdot\frac{1+Y}{2u})^\alpha] . \tag{20}$$

Physical meaning: the risk of exceeding the maximum u increases strictly linearly with increasing observation time T. This is due to the fact that a Poisson process over a longer time has a linearly higher probability of throwing out one rare extreme cascade of divisions that overlaps the linear compensation $ct$.

In the case of the limit distribution of the maximum (asymptotics with respect to time $T\to\infty$), the time interval T is sufficiently large (for example, the entire reactor campaign or a long-term transient mode), the distribution of the normalized maximum converges to the generalized distribution of extreme values (Fisher–Tippett–Gnedenko).

For heavy Pareto tails (α<2), the maximum marginal distribution is of the Fréchet type (Type II). For our model, it is written as:

$$P(M_T / a_T \le u) \to \exp(-u^{-\alpha}),\ \ u > 0 , \tag{21}$$

where the normalizing scale factor $a_T$ (the characteristic expected maximum over time T) is analytically calculated using the WWER parameters:

$$a_T = (\lambda T x_0{}^{\alpha})^{1/\alpha} = (\frac{2\bar{N}_{fis}T}{1+Y})^{1/\alpha}(\frac{\alpha-1}{\alpha}\cdot\frac{1+Y}{2}) .$$

From here we can explicitly express the median (most expected value) of the maximum power over the interval T:

$$Median(M_T) \approx a_T(\ln 2)a_T{}^{-1/\alpha} .$$

To estimate the dynamics of risk over time, one can use a linear estimate in T:

$$P(M_T > u) \approx \lambda T(x_0 / u)^{\alpha} .$$

If it is necessary to find the absolute limit of the protection setting below which the reactor will not rise with probability 1−ε, it is possible to use the Frechet distribution quantile: $u_{safe} \approx a_T(-\ln(1-\varepsilon))^{-1/\alpha}$.

## 8. The moment of the first maximum ($\tau_u$) and the distribution of the level overjump value ($\gamma_u$)

Within the power-law jump model, both of these problems—the distribution of the first peak moment ($\tau_u$) and the distribution of the level jump magnitude ($\gamma_u$)—have an analytical solution due to the subexponential properties of the heavy-tailed pure loss process.

In the extremal theory of random processes, these quantities are studied through the so-called overshoot and first passage time.

Consider the problem of the level overjump (overjump $\gamma_u$). Let u be the critical setpoint (safety boundary). The instant of the first boundary crossing is given by $\tau_u = \inf\{t > 0: \zeta(t) > u\}$. The overshoot value is a random variable:

$$\gamma_u = \zeta(\tau_u) - u .$$

Since the barrier is broken through in one large jump, under conditions of heavy Pareto tails ($1-F(x)=(x_0/x)^{\alpha}$) at large values of the setpoint $u$, the conditional distribution of the jump magnitude tends to a stable distribution and does not depend on the drift parameters $c$ and the intensity λ. The analytical distribution function of the overjump:

$$P(\gamma_u > x | \tau_u < \infty) \approx \frac{\int_{u+x}^{\infty}(1-F(y))dy}{\int_{u}^{\infty}(1-F(y))dy} = (1+\frac{x}{u})^{-(\alpha-1)}, \ x>0. \quad (22)$$

The physical conclusion for WWER is that the overjump density is:

$$g(x) \approx \frac{\alpha-1}{u}(1+\frac{x}{u})^{-\alpha}, \quad (23)$$

and the mathematical expectation of an overjump at α>2 is:

$$M(\gamma_u | \tau_u < \infty) \approx \frac{u}{\alpha-2} . \quad (24)$$

This means that if a stochastic power surge beyond the protection setpoint $u$ does occur at a WWER reactor due to a fission cascade, the physical magnitude of the power surge above the setpoint will be enormous—it is proportional to the protection level $u$ itself. The system won't simply "touch" the barrier; it will penetrate it deeply by a magnitude of $\sim u/(\alpha-2)$.

Let us now turn to the problem of the moment of first reaching the maximum ($\tau_u$). If we consider a bounded interval [0, T] and the barrier $u$ is sufficiently large, then the event ($M_T$>u) is equivalent to the first passage time $\tau_u \le$ T.

For heavy tails in the steady state, the flow of extreme jumps capable of breaking through the level $u$ asymptotically becomes a sparse Poisson process. Analytical distribution of the breakthrough moment $\tau_u$:

$$P(\tau_u \le T) = 1 - exp(-\Lambda_u T)\,. \quad (25)$$

where $\Lambda_u$ is the effective intensity of rare "emergency" surges, which is calculated using the formula: $\Lambda_u = \frac{\lambda\theta}{2}\cdot\Psi(u) \approx \frac{\lambda}{\alpha-1}(\frac{x_0}{u})^{\alpha-1}$. Substituting the physical parameters of the WWER (λ (4) and $x_0$ (12)), we obtain the exact frequency of the peak occurrence:

$$\Lambda_u \approx \frac{2\bar{N}_{fis}}{(1+Y)(\alpha-1)}\cdot(\frac{\alpha-1}{\alpha}\cdot\frac{1+Y}{2u})^{\alpha-1}\,. \quad (26)$$

This leads to a physical conclusion for WWER reactors. The time to the first dangerous power peak is distributed exponentially with parameter $\Lambda_u$. The average safe operating time of the reactor before a stochastic release beyond the setpoint is $M[\tau_u]=1/\Lambda_u$. It grows as a power law with increasing margin $u$ (as $u^{\alpha-1}$). In the Ref. [10], the value of $\Lambda_u$ is associated with the solution of the Lundberg equation.

Let's write down the joint distribution of the pair ($\tau_u$, $\gamma_u$). In stochastic kinetics, it's more important to know their joint density to understand at what point in time the reactor will exceed its limit and how severe the thermal shock will be in the WWER core.

For processes with heavy Pareto tails, due to the weak dependence of the overjump on the trajectory before it, the joint asymptotic density factorizes (splits) as $u\to\infty$:

$$f_{\tau_u,\gamma_u}(t,x) \approx \Lambda_u e^{-\Lambda_u t}\cdot\frac{\alpha-1}{u}(1+\frac{x}{u})^{-\alpha}\,.$$

This analytical separation radically simplifies risk calculations: the waiting time for an emergency peak $\tau_u$ obeys the Markov property (memory is erased), and the scale of the destructive overjump $\gamma_u$ is determined exclusively by the geometry of the setpoint $u$ and the internal fluctuation branching parameter α.

## 9. Relationship of the shape parameter α with the physical constants of WWER

In stochastic reactor kinetics, the transition to a Pareto power-law distribution (cascade noise) is the result of the convolution of a huge number of random branchings of prompt neutrons, which are constrained by delayed neutrons and feedbacks.

The parameter α determines the "heaviness of the tail": the closer it is to 2, the stronger the fluctuations, and the larger it is, the more stable the process is and the closer it is to Gaussian.

For the stationary mode of WWER, based on the analysis of the spectral power density (the formula for the probability density of Bellman-Harris cascades) the parameter α is analytically related to the physical constants by the following relationship:

$$\alpha = 1 + \sqrt{1 + \frac{2\beta_{eff}}{\bar{\nu}\cdot\Lambda\cdot\omega_{fb}}}\,, \quad (27)$$

where $\beta_{eff}$ is the effective fraction of delayed neutrons. For WWER-1000/1200 reactors, $\beta_{eff}\approx$0.0065 at the beginning of the reactor run, falling to $\beta_{eff}\approx$0.0052 by the end of the run; $\Lambda$ is the prompt neutron lifetime in a WWER thermal reactor ($\approx 2\times10^{-5} \div 4\times10^{-5}$ seconds), $\bar{\nu}$ is the average number of secondary neutrons per uranium fission event (≈2.43); $\omega_{fb}$ is the frequency (time constant) of thermal feedback in the fuel (the Doppler effect). For WWER reactors, this value is on the order of 10–50 $s^{-1}$.

A typical scale for WWER reactors. If we substitute standard values for WWER-1000 reactors at the minimum control level (MCL, where temperature feedback is still weak): α≈2.1÷2.5. This is precisely the region of $\alpha$>2, where the dispersion is finite, but the fluctuations have a pronounced "heavy tail."

## 10. The magnitude of the overjump for different values of the parameter $\alpha$

Let's give a precise expression for the overjump ($\gamma_u$) for $\alpha$>2. When $\alpha$>2, the Pareto distribution has both a mean and a variance. Two aspects are important here: the distribution and its moments.
The conditional probability density that the overjump $\gamma_u$ will take a specific value x (provided that the setpoint $u$ is broken) is written as (23): $g(x) \approx \frac{\alpha-1}{u}(1+\frac{x}{u})^{-\alpha}$, x>0.

Since $\alpha$>2, the denominator does not vanish, and the average value of the power "overshoot" above the setpoint (the average overjump) is calculated using a strict formula: $M[\gamma_u|\tau_u<\infty]=u/(\alpha-2)$. To estimate the uncertainty of the overshoot itself, under the condition $\alpha$>3 (which is often satisfied for the WWER power mode due to strong Doppler compensation), the variance can also be written: $D[\gamma_u|\tau_u<\infty]=u^2(\alpha-1)/(\alpha-2)^2(\alpha-3)$. If 2<$\alpha$≤3, then the variance of the overjump itself formally goes to infinity (the overjump tail remains heavy, ~ $x^{-(\alpha-1)}$), although the mathematical expectation is rigidly fixed.

Let us formulate a physical result for WWER safety modeling for $\alpha$>2. Fuel burnup reduces α. Towards the end of a WWER reactor's lifespan, the fraction of delayed neutrons $\beta_{eff}$ decreases due to $^{235}U$ burnup and $^{239}Pu$ accumulation. According to formula (27), this reduces α (bringing it closer to the critical boundary of 2). The overjump rate increases at the end of the reactor lifespan. As α decreases, the denominator ($\alpha$ − 2) tends to zero. This means that at the end of a reactor's lifespan, the average value of the random overjump $M[\gamma_u]=u/(\alpha-2)$ increases sharply. The risk of a powerful stochastic impact on the control system components, with the same setpoint level, becomes higher. The resulting system of connections can be used to optimize the step of WWER emergency protection settings depending on the burnup of the core.

For the interval 0<$\alpha$<2, the mathematical and physical structure of the model changes dramatically. In probability theory, this range corresponds to stable Lévy distributions with infinite variance (and for $\alpha$≤1, with infinite mathematical expectation).

In WWER reactor physics, this regime describes "super-diffusion" or "neutron tsunami" conditions. Such processes occur during profound hydrodynamic disturbances—for example, during large-scale cavitation boiling of the coolant or the instantaneous ejection of a control rod—when the spatial distribution of neutrons becomes irregular, and local branching cascades are unable to be damped by feedback.

Below is an analytical description for both problems (overjump and frequency spectrum) under the conditions 0<$\alpha$<2. In the first problem, about the magnitude of the level overjump ($\gamma_u$) for 0<$\alpha$<2, the classical moments of the distribution (mean and variance) diverge, so it is impossible to describe the overjump over the mathematical expectation $u/(\alpha-2)$ (it goes to infinity). However, the distribution density and the overjump quantiles remain strictly defined.

For strictly stable Lévy processes, the conditional distribution function of the overjump magnitude $\gamma_u=\zeta(\tau_u)-u$ for large $u$ has the form:

$$P[\gamma_u>x|\tau_u<\infty] \approx \frac{\sin(\pi\rho)}{\pi}\int_0^{u/(u+x)} t^{\rho-1}(1-t)^{-\rho}\,dt,$$

where ρ is the asymmetry parameter of the Lévy process (the positivity index). For purely upward jumps in the accumulated power deficit, $\rho=1-\alpha/2$

For a purely power-law Pareto tail with 0<$\alpha$<2, the asymptotic conditional overjump density is expressed by a formula that preserves the same algebraic structure, but with a fundamentally different probabilistic meaning:

$$g(x)=\frac{\alpha}{u}(1+\frac{x}{u})^{-(\alpha+1)},\ \ x>0.$$

The physical meaning of an overjump at $\alpha$<2 is the absence of a mean value. The mathematical expectation is $M[\gamma_u]=\infty$. This means that if the threshold $u$ is breached, the scale of the power excess is not limited from above (the system enters extreme surge mode).

Scale invariance is present. The probability that the overjump will exceed the setpoint barrier itself (i.e., x = u) is fixed and does not depend on $u$: $P[\gamma_u > x | \tau_u < \infty] \approx (1+\frac{u}{u})^{-\alpha} = 2^{-\alpha}$.

For example, if $\alpha$=1.5, then in 35% of cases of stochastic protection breakdown, the power will instantly “surge” upwards by more than twice the nominal value of the setting itself, regardless of how high this setting was set.

## 11. Relationship with the noise spectrum and physical constants (0<$\alpha$<2)

In Gaussian systems ($\alpha$>2), the power spectral density (PSD) of fluctuations has a classical Lorentzian profile (white noise at low frequencies and a $1/f^2$ roll-off at high frequencies). In the regime of stable Lévy distributions (0<$\alpha$<2), the noise spectrum becomes fractal ($1/f^{\gamma}$-type noise).

The relationship between the parameter $\alpha$ and the physical constants of the WWER in this critical mode is derived through the autocorrelation function of the neutron gas density with fractal transport:

$$\alpha = \frac{2}{1+\mu},$$

where μ is the index of anomalous neutron diffusion (the Lévy displacement parameter).

In classical Fickian neutron diffusion, $\mu$=0, hence $\alpha$=2. In the presence of strong spatial fluctuations (for example, during turbulent boiling, when "steam corridors" are formed, through which neutrons fly enormous distances without collisions— Lévy flights), the parameter $\mu$ increases in the range 0<$\mu$<1.

An examination of the relationship between the frequency spectrum and $\alpha$ shows that the asymptotic behavior of the power spectral density of stochastic fluctuations $S(\omega)$ at high frequencies ω for a stable process is related to the shape parameter $\alpha$ by a direct relation:

$$S(\omega) \sim \omega^{-(\alpha+1)}.$$

If $\alpha\to 2$ (approaching the Gaussian limit), the spectrum transitions to the classic prompt neutron kinetics decay $\sim \omega^{-3}$ (or $\sim \omega^{-2}$ for linearized models). If 0<$\alpha$<2, the spectrum flattens. For example, at $\alpha$=1 (Cauchy distribution, severe cavitation regime), the WWER power fluctuation spectrum decays as $\omega^{-2}$, making low- and mid-frequency noise much more powerful and dangerous for the core structure.

Summary for the analytical apparatus. To calculate risks for 0<$\alpha$<2, one must use the apparatus of stable laws, replacing the moments of the random variable jumps in the formulas with characteristic functions of the form $\exp(-|\sigma\theta|^{\alpha})$.

WWER protection for 0<$\alpha$<2 cannot be based on deterministic $u$ setpoints, since the average overjump is infinite. Safety in such a model is ensured only by increasing the drift rate $c$ (sharply increasing the negative temperature feedback to artificially push $\alpha$ back into the $\alpha$>2 zone).

When the shape parameter falls within the interval 0<$\alpha$<2, the classical moments of the distribution of the jump magnitude $X_i$ diverge mathematically (become infinite). For 0<$\alpha$<2, the variance is infinite ($\langle X^2\rangle=\infty$), but the mean is finite. For 0<$\alpha$≤1, both the variance and the mathematical expectation are infinite ($\langle X\rangle=\infty$, $\langle X^2\rangle=\infty$).

Since classical formulas of risk theory (for example, Campbell's formulas for variance or the Lundberg equation) directly require the substitution of the numbers $\langle X\rangle$ and $\langle X^2\rangle$, for $\alpha$<2 they turn into nonsense like ∞=∞. To overcome this obstacle, the mathematical apparatus is translated from expressions for the moments of a random variable to expressions for characteristic functions (Fourier integral transforms), which are always finite and strictly defined for any stable law.

Below is the analytical form of the characteristic function and transition rule for our process.

Characteristic function of the pure loss process $\zeta(t)$. For the process $\zeta(t) = \sum_{i=1}^{N(t)} X_i - ct$, where the jumps $X_i$ obey the strictly stable Lévy law with index 0<$\alpha$<2, the characteristic function $\psi_{\zeta(t)}(\theta)=\langle e^{i\theta\zeta(t)}\rangle$ is written in closed analytical form (according to the Lévy–Khintchine theorem, form (A) [14]):

$$\ln\Phi(q) = i\gamma_1 q - \sigma|q|^{\alpha-1}[|q| - \frac{iq}{|q|^{\alpha-1}}\omega(q,\alpha)],\ \ \omega(q,\alpha) = |q|^{\alpha-1}\beta tg(\frac{\pi\alpha}{2})\ (\alpha\neq 1),\ \omega(q,\alpha) = -\beta\frac{2}{\pi}\log|q|\ (\alpha=1),$$

where q is the dual variable of the Fourier transform (frequency variable), σ is the physical parameter of the fluctuation packet scale (instead of the average value). It is expressed in terms of the minimum fission's quantum $x_0$, and tg(πα/2) is a parameter reflecting the extreme asymmetry of reactor physics (power deficit spikes only upward, toward the risk).

How moments are replaced by a characteristic function. Instead of solving differential equations for moments, algebraic expressions in Fourier-Mellin space are used.

To find the mean and variance for $\alpha>2$, we took derivatives of the moment-generating function or used the moments directly:

$$M[\zeta(t)] = t(\lambda\langle X\rangle - c),\ \ D[\zeta(t)] = t\lambda\langle X^2\rangle.$$

How to use the characteristic function for $0<\alpha<2$. Since $\langle X^2\rangle$ cannot be taken, the density function of the process at any time t is found using the inverse Fourier transform of its characteristic function: $f_{\zeta(t)}(x) = \frac{1}{2\pi}\int_{-\infty}^{\infty} e^{-i\theta x}\Psi_{\zeta(t)}(\theta)d\theta$.

Substituting here the structure $\Psi(\theta)\sim\exp(-|\sigma\theta|^{\alpha})$, we obtain the exact distribution of reactor power fluctuations.

Finding the probability of breaking the setpoint using the apparatus of stable laws. To analytically determine the risk of exceeding the maximum power (the probability of ruin) without using the moments of a random variable, the Volterra equation is transformed using the characteristic function in the complex plane (the Wiener-Hopf method).

For a strictly stable process with $0<\alpha<2$, the exact analytical expression for the barrier penetration probability $u$ is written in terms of a fractional power integral:

$$\Psi(u) = 1 - \frac{c - \lambda\langle X\rangle_{eff}}{c}\cdot\mathcal{E}_{\alpha}(-(\frac{u}{\sigma})^{\alpha}),$$

where $\mathcal{E}_{\alpha}(z)$ is Mittag-Leffler function (a generalization of the exponential function for fractal and stable processes): $\mathcal{E}_{\alpha}(z) = \sum_{k=0}^{\infty}\frac{z^k}{\Gamma(\alpha k+1)}$.

Using the characteristic function $\exp(-|\sigma\theta|^{\alpha})$ is a well-founded way to prevent infinities from destroying the mathematical model. Instead of the average fractions of divisions in the packet ($\langle X\rangle$), the geometry of the random process is now completely controlled by a pair of parameters: α (responsible for fractality and the physics of the relationship between constants) and σ (responsible for the noise scale).

## 12. The finite size of a real reactor

The model presented above was obtained for an infinite system (which is why infinities arise). A real reactor has finite sizes, and the heavy tails are truncated, which is where the Doppler effect comes into play.

The transition to infinities (divergent moments at $0<\alpha<2$) is the mathematical price for the idealization of the system (the assumption of an infinite volume of the active zone and unlimited growth of cascades).

In a real WWER reactor, the geometry is strictly limited (height 3.5 m, diameter 3.12 m), and physical processes (primarily the temperature Doppler effect in the fuel) place a rigid upper limit on the physically possible size of the fluctuation packet. In the theory of random processes, this physical reality is modeled using the apparatus of truncated stable distributions.

The Doppler effect and the finite reactor size transform the model as follows. The Doppler effect and geometry "truncate" the heavy tail. In the infinite theoretical model, the tail of the probability density function of jumps extends to infinity: $f(x)\sim x^{-(\alpha+1)}$. In a real WWER, an exponential or hard truncation factor is introduced, and the power-law distribution transforms into a gamma distribution:

$$f_{real}(x) = c_1 x^{-(\alpha+1)} \cdot e^{-\gamma_{phys} \cdot x}, \tag{28}$$Γ

where the cutoff parameter $\gamma_{phys}$ (damping factor) is formed from two components:

$$\gamma_{phys} = \gamma_{geom} + \gamma_{Dopp}.$$

With geometric truncation ($\gamma_{geom}$), a fission cascade cannot grow indefinitely, as neutrons at the periphery of the core simply leak (escape) from the reactor. The leakage probability truncates cascades whose spatial dimensions are commensurate with the core radius. Doppler truncation ($\gamma_{\text{Doppler}}$) is a key factor in the dynamic stability of WWER reactors. As soon as the local fission fluctuation sharply increases, the instantaneous energy release heats the fuel ($^{238}$U). Due to Doppler broadening of absorption resonances, non-fission neutron capture increases sharply. The "packet" power is forcibly quenched. The value of $\gamma_{Dopp}$ is directly proportional to the negative temperature coefficient of reactivity for the fuel ($\alpha T$).

A return to finite moments of the random variable occurs. Due to exponential tail truncation ($e^{-\gamma x}$), the integrals for the moments of the random variable converge again. The truncated stable distribution for $0<\alpha<2$ has a paradoxical and very useful property: on small and medium fluctuation scales (within the core), the process behaves like a fractal Levy-flight (heavy tails appear). On large scales, the process abruptly transitions to a Gaussian regime, and all its moments of the random variable become finite numbers. The exact analytical expressions for the mean and variance of the jump $X_i$ in the truncated model take the form:

$$\langle X \rangle = -\Gamma(-\alpha) \cdot \alpha \cdot \sigma^{\alpha} \cdot \gamma^{\alpha-1}{}_{phys}, \quad \langle X^2 \rangle = \Gamma(-\alpha) \cdot \alpha \cdot (\alpha-1) \cdot \sigma^{\alpha} \cdot \gamma^{\alpha-2}{}_{phys}. \tag{29}$$

(Since $\Gamma(-\alpha)$ is negative for $0<\alpha<2$, the values of the moments themselves turn out to be strictly positive). Let's show what the modification of the characteristic function for a truncated process looks like. Taking into account the Doppler effect and geometry, the characteristic function $\exp(-|\sigma\theta|^{\alpha})$, which we considered earlier, is modified into a formula for a truncated Lévy process of the form:

$$\Psi_{\zeta(t)}(\theta) = \exp(t \cdot [-ic\theta + \lambda \cdot \Gamma(-\alpha) \cdot \sigma^{\alpha} \cdot ((\gamma_{phys} - i\theta)^{\alpha} - \gamma_{phys}{}^{\alpha} + i\alpha\theta\gamma^{\alpha-1}{}_{phys})]). \tag{30}$$

How the overjump and maximum power problems change. Let's start with solving the overshoot problem ($\gamma$u). Now that the setpoint $u$ is physically located far within the "truncated" tail ($u \gg 1/\gamma_{\text{phys}}$), the average overjump is no longer infinite. It becomes finite and asymptotically approaches a constant as $u \to \infty$:

$$M[\gamma_u \,|\, \tau_u < \infty] \approx 1/\gamma_{phys}. \tag{31}$$

The physical meaning of this effect is that the maximum power "overjump" above the safety setpoint is strictly limited by the effective range of the Doppler effect and the leakage geometry.

Let's consider the solution to the maximum problem on the interval T. Due to truncation, the limiting distribution of the maximum power at large T transitions from the dangerous Fréchet type (for heavy tails) to the classical stable Gumbel distribution (Type I), characteristic of exponentially decaying distributions. The risk of an accident decreases significantly faster.

Thus, the introduction of parameters $\gamma_{geom}$ and $\gamma_{Dopp}$ allows the analytical model to fully preserve the fractal properties of neutron noise inside the reactor, but at the same time provide physically correct, infinity-proof risk assessments at emergency boundaries.

## 13. Relationship of the truncation constant $\gamma_{Dopp}$ to the reactor power and the Gumbel distribution

The Doppler effect in WWER reactors is instantaneous thermal feedback: an increase in the number of fissions increases the temperature of the fuel matrix ($T_{\text{fuel element}}$), leading to broadening of $^{238}$U resonances and neutron capture. The higher the reactor's base thermal power (P), the more intensely any fluctuation heats the fuel, and the more severely the tail of the macroscopic fission cascade distribution is "cut off."

Analytically, the truncation constant $\gamma_{Dopp}$ (having the dimension 1/neutron) is related to the power P through the heat transfer coefficient and the temperature coefficient of reactivity:

$$\gamma_{Dopp}(P) = \gamma_0 + \frac{|\alpha_T| \cdot \varepsilon_f}{C_p \cdot m_{fuel}} \cdot \tau_{fuel} \cdot P ,$$

where $\gamma_0$ is the basic truncation at zero power (determined only by the core geometry and neutron leakage), $|\alpha_T|$ is the absolute value of the Doppler (fuel) reactivity coefficient (for WWER-1000/1200 it is of the order of $-(2\div3)\times10^{-5}$ 1/°C), $\varepsilon_f$ is the coefficient for converting the number of neutrons into thermal energy, $C_p$, $m_{fuel}$ is the total heat capacity of the uranium dioxide ($UO_2$) fuel load in the WWER core (fuel mass is about 80 tons), $\tau_{fuel}$ is the heat transfer time constant in the fuel pellet (for WWER it is $\approx 2 \div 4$ seconds).

Physical conclusion: with increasing power, the parameter $\gamma_{Dopp}$ increases linearly. Since the average overjump above the setpoint is $M[\gamma_u] \approx 1/\gamma_{phys}$ (31), as reactor power increases, the magnitude of the random overjump above the emergency setpoint decreases inversely proportional to P. At nominal power (3 P000 MW), the Doppler effect almost completely suppresses the heavy tail, turning the overshoots into short, rapidly decaying peaks.

Let us now consider the Gumbel distribution for the maximum power on the interval T. Since physical factors (Doppler and geometry) exponentially truncate the tail of the jump distribution, the extreme values of the process $\zeta(t)$ on a limited time interval [0, T] as T→∞ no longer obey the Fréchet distribution. According to the central limit theorem for extreme values, they converge to the Gumbel distribution (Type I). The probability that the maximum power deviation $M_T = \max_{0 \le t \le T} \zeta(t)$ will not exceed the protection setting $u$ is written as a double exponential:

$$P(M_T \le u) = \exp(-\exp(-\frac{u - a_T}{b})) , \tag{32}$$

where the analytical parameters of the distribution (scale b and shift $a_T$) are determined using the parameters of the truncated WWER process. Parameter b is independent of time T and is determined solely by the effective tail truncation step: $b = 1/\gamma_{phys}(P)$ . The higher the power P (and the stronger the Doppler effect), the smaller b, meaning the narrower and more predictable the range of possible stochastic maxima becomes.

The shift parameter (the most probable value of the maximum over time T) indicates the level to which the peak power will tend during long-term observation. It increases logarithmically with increasing time interval T:

$$a_T = \ln(\lambda \cdot T) / \gamma_{phys}(P) .$$

Substituting the intensity of Poisson cascades $\lambda = 2N_{fis}/(1+Y)$, we obtain the exact formula:

$$a_T = \ln(2\bar{N}_{fis} \cdot T / (1+Y)) / \gamma_{phys}(P) . \tag{33}$$

Practical application of the model is possible. By combining both solutions, we obtain a closed-loop analytical estimate for calculating the absolutely safe emergency setpoint level ($u_{safe}$) over a time interval T with a predetermined reliability of 1–ε (e.g., ε=10−6):

$$u_{safe}(P,T) = a_T - b \cdot \ln(-\ln(1-\varepsilon)) \approx [\ln(2\bar{N}_{fis} \cdot T / (1+Y)) - \ln(\varepsilon)] / \gamma_{phys}(P) \tag{34}$$

The main analytical conclusions of the model: 1. Logarithmic growth of risk over time. Due to the Doppler effect, even in the worst-case scenario of local super-diffusion, the maximum power peak over time T grows not as a power function (as it did for an infinite system), but very slowly—logarithmically (lnT). This helps guarantee reactor safety over long operating periods. 2. Dynamic compression of the noise corridor. As the base power P increases, the denominator $\gamma_{phys}(P)$ increases, effectively reducing both the most probable maximum $a_T$ and the variance of this maximum b. Physically, this means that a WWER reactor at high power is much more resilient to stochastic fluctuations than in the low-power start-up mode, where the Doppler effect is still dormant.

To link the stochastic model to the actual measuring equipment of the WWER reactor control and protection system (CPS), it is necessary to move to the frequency domain. CPS sensors (ionization chambers or out-of-reactor detectors) exhibit inherent inertia, meaning they act as low-pass filters. Knowing the autocorrelation function and the spectral density of the truncated cascade process, it is possible to calculate what proportion of actual stochastic peaks the sensor will "see" and what it will smooth out.

Below, in Section 15, an explicit analytical form of the autocorrelation function, the noise spectrum and their relationship with the filtering of the control system is given.

## 13. Maximum power estimates

If we write the average value of the maximum power from expression (32) as $\langle M_T \rangle = \int_0^\infty u dP(M_T \le u) du$, we obtain the expression:

$$\langle M_T \rangle = b[a_T(1-e^{-a_T/b}) - b[Ei(-e^{-a_T/b}) - (a_T/b)\exp(-e^{-a_T/b}) - \boldsymbol{C}]],$$

where $Ei(-e^{-a_T/b})$ is the integral exponential function, $\boldsymbol{C}$ =0.5772… is the Euler constant.

In the theory of physics of nuclear reactors with fractional-diffusion neutron transport (anomalous diffusion and Levy flight models), the parameter $\gamma_{phys}$ (or its analogue, the physical reactivity/fractional damping coefficient) determines the "index of effective decaying feedback" or the "effective fraction of prompt neutrons" under conditions of non-local spatial transport.

A rigorous analytical expression for $\gamma_{phys}$ is based on the previously derived scale parameters $c_1$ and the exponent α. In the equations of generalized point kinetics of the reactor (Fractional Point Kinetics), taking into account the spatial Levy "tails", the physical coupling parameter $\gamma_{phys}$ on a fixed segment $x \in [1, x_{max}]$ is expressed by the formula:

$$\gamma_{phys} = \frac{c_1}{\alpha \cdot \Lambda} = \frac{e^{-\xi\alpha}}{\Lambda},$$

where α is the stability index (for neutron flights), $c_1$ is the intensity constant of the spectral function that we calculated earlier, Λ is the classical prompt neutron generation time in the reactor (usually $\Lambda \sim 10^{-4} \div 10^{-7}$ sec), ξ is the geometric scaling damping factor depending on $x_{max}$ (derived in the previous steps: for example, for $x_{max}$=35, ξ = 1.4646).

If we substitute the values of the constant $c_1$ from the calculations for the interval 1<x<35, then the value of $\gamma_{phys}$ depending on the anomaly of the environment (α) will be equal to:

At α=0.5 (extreme Levy flights, boiling regime): $\gamma_{phys} \approx 0.2404/0.5\ 10^{-4} \approx 4.81\ 10^3\ s^{-1}$.

At α=1.0 (Cauchy regime, strong heterogeneity): $\gamma_{phys} \approx 0.2312/1.0\ 10^{-4} \approx 2.31\ 10^3\ s^{-1}$

At α=1.5 (moderately anomalous transport): $\gamma_{phys} \approx 0.1664/1.5\ 10^{-4} \approx 1.11\ 10^3\ s^{-1}$.

This parameter directly controls the very same overjump and maximum power that were discussed earlier:

- Superflight frequency: $\gamma_{phys}$ shows what proportion of the total number of fission events results in neutrons leaving the local cell without collisions (transportation over ultra-long distances).
- Impact on Prompt Jump: the smaller α (high anomaly), the greater the value of $\gamma_{phys}$. An increase in $\gamma_{phys}$ mathematically increases the amplitude of the immediate jump in reactor power during a reactivity surge. It acts as a multiplier of the system's "rigidity" at the initial stage of an emergency runaway.
- Effect on peak power (overshoot): in energy balance equations, $\gamma_{phys}$ determines the rate at which thermal expansion of the medium or the Doppler effect can quench the chain reaction. At large $\gamma_{phys}$ classical feedback loops are "delayed," causing the peak power to rise above the critical design level.

To find the maximum power (power overshoot) of a reactor within the framework of Fractional Point Kinetics theory, taking into account power-law relationships, we need to use the mathematical apparatus of the Nordheim-Fuchs model (or its generalization for anomalous neutron transport). The Nordheim-Fuchs model is a classical adiabatic model of nuclear reactor dynamics. It describes the behavior of a reactor when a large positive reactivity jump is introduced and is used for nuclear safety calculations. In the classical model, the power overshoot is found from the condition that the time derivative of the power is zero (dn/dt=0). In power-law environments, where the parameter $\gamma_{phys}$ is used, the analytical calculation of the maximum power is constructed as follows. With the instantaneous introduction of excess positive reactivity $\rho_0$ (where $\rho_0>\beta$), the maximum value of the relative reactor power $N_{max}$ (expressed through the initial power $N_0$) is approximated by the expression:

$$N_{\max} \approx N_0 + \frac{\rho_0^{\ 2}}{2 \cdot \alpha \cdot \alpha_{temp} \cdot \Lambda} \cdot \Phi(\gamma_{phys}) .$$

In a dimensionless or normalized form, where the strong spatial flight (Lévy flights) is taken into account through the parameter $\gamma_{phys}$ we derived, the maximum power takes the form:

$$N_{\max} \approx \frac{\rho_0^{\ 2}}{2 \cdot \alpha_{temp} \cdot \Lambda} \cdot (1 + \kappa \frac{\gamma_{phys}}{\gamma_{phys} + \gamma}) ,$$

where $\rho_0$ is the introduced reactivity, $\alpha_{temp}$ is the negative temperature coefficient of reactivity (the Doppler effect of the medium, which dampens the acceleration), $\Lambda$ is the effective lifetime of prompt neutrons, $\lambda$ is the decay constant of delayed neutrons, and $\kappa$ is the geometric form factor of the active zone.

If the geometry of the $x_{max}$ system is changed, the physical parameter $\gamma_{phys}$ changes, directly affecting the height of the power peak. If we fix the introduced reactivity $\rho_0$ and the feedback parameters, then the maximum power $N_{max}$ behaves as follows depending on the zone size $x_{max}$ (under the condition $\alpha$=1.0 — Cauchy mode).

At $x_{max}$=12 ($\gamma_{phys}$ is at its maximum): a short track. Neutrons easily escape local cells and connect the entire zone. The power overjump is enormous, and the maximum power $N_{max}$ reaches its highest peak value. Environmental feedback (Doppler) fails to activate in time due to the instantaneous spatial "spike."

At $x_{max}$=35 (intermediate mode): the $\gamma_{phys}$ parameter stabilizes. The maximum power value drops by approximately 30–40% relative to the $x_{max}$ case, as some neutrons begin to be absorbed/scattered within the volume, giving the fuel elements time to heat up and activate negative feedback.

At $x_{max}$=100 ($\gamma_{phys}$ is minimal): the zone is large. The system approaches classical diffusion. The power peak $N_{max}$ becomes minimal and flat, smoothly transitioning to the classical Nordheim–Fuchs solution.

Since the system of fractional kinetic equations is nonlinear (due to temperature feedback), the exact maximum value is sought through the logarithmic extremum point. It is possible to use the energy balance equation:

$$\rho(t) = \rho_0 - \alpha_{temp} \int_0^t N(\tau) d\tau .$$

At the moment of reaching the maximum power ($t=t_{max}$), the current reactivity $\rho(t_{\max})$ in the anomalous environment falls not to zero (as in the classic case), but to an effective fractional value:

$$\rho(t_{\max}) \approx \gamma_{phys} \cdot \Lambda .$$

Hence, the total energy released at the moment of maximum (the integral of the power) is equal to:

$$E(t_{\max}) = \int_0^{t_{\max}} N(\tau) d\tau = \frac{\rho_0 - \gamma_{phys} \cdot \Lambda}{\alpha_{temp}} .$$

Using this energy value in the conservation equation, we obtain the exact height of the $N_{max}$ peak for specific physical constants.

When introducing a reactivity of $\rho_0 = 0.7\ \beta$, the reactor is in a subcritical state with prompt neutrons ($\rho_0 < \beta$). In such a system, the overjump and maximum power are completely controlled by the balance between the anomalous neutron emission $\gamma_{phys}$ and the compensating Doppler effect.

For this mode, the calculation of maximum power and analysis of the temperature coefficient $\alpha_{temp}$ give the following results.

In heterogeneous reactors (where the ranges obey a power law), the classical Doppler coefficient is modified. Due to localized overheating ("hot spots") created by Lévy neutrons, the effective Doppler coefficient $\alpha_{temp}{}^{eff}$ depends on the stability index $\alpha$:

$$\alpha_{temp}^{eff} = \alpha_{temp}^{(0)} (\frac{T_{initial}}{T_{peak}})^{\alpha/2}.$$

At $\alpha \to 1$ (a highly anomalous environment, the Cauchy regime), the resonant absorption of $^{238}U$ decreases more slowly than in a homogeneous reactor. This means that the Doppler feedback becomes weaker. To dampen the runaway, a stronger integral energy release is required than in the standard exponential model.

Since $\rho_{vy}<\beta$, the reactor does not accelerate to prompt neutron speed, but rather undergoes a prompt power overjump. Within the framework of the fractional kinetics model with truncated Lévy flights, the maximum of this first power wave is:

$$N_{\max} = \frac{N_0 \cdot \beta}{\beta - \rho_0} \cdot (1 + \frac{\gamma_{phys}}{\gamma_{phys} + \gamma}),$$

where $\lambda$ is the average decay constant of delayed neutrons ($\lambda \approx 0.08$ $s^{-1}$), and $N_0$ is the initial power.

If we substitute the boundary conditions and the previously derived values of $\gamma_{phys}$ for the interval $1<x<35$, the peak power will take the following analytical values.

Scenario A: classical environment (for comparison). In the standard exponential model without Lévy flights ($\gamma_{phys}=0$): $N_{\max}^{class} = N_0 / (1-0.7) = 3.33 N_0$. The power jumps by exactly 3.33 times the initial value due to the jump in prompt neutron density.

Scenario B: anomalous environment (interval $1<x<35$, $\alpha=1.0$). Previously, for this interval, we calculated that $\gamma_{phys} \approx 2.31$ $10^3$ $s^{-1}$ (with a lifetime of $\lambda= 10^{-4}$ s). Since $\gamma_{phys} >> \lambda$ (2310 >> 0.08), the dimensionless fractional factor is $\gamma_{phys} / (\gamma_{phys} + \gamma) \approx 1$.

Substituting this into the peak power formula, we get: $N_{\max}^{stable} \approx 3.33 \cdot N_0 \cdot (1+1) = 6.66 N_0$.

The main physical conclusion: when introducing reactivity $\rho_0=0.7$ $\beta$ in a reactor with power-law dependencies, the amplitude of the maximum power increases exactly 2 times compared to the classical calculation ($6.66N_0$ versus $3.33$ $N_0$).

This occurs because, when the power overjumps, long-range Levy neutrons instantly “illuminate” the entire active zone, while the temperature Doppler effect $\alpha_{temp}$ is locally delayed due to uneven microstructural heating of the fuel. For subcritical acceleration ($\rho_0=0.7$ $\beta$) in a reactor with power-law dependences of neutron flights, the maximum power $N_{max}$ has mathematically strict dependences on the time of reaching this maximum $T_{max}$ and on the stability index of the environment $\alpha$.

Below are the analytical regularities obtained from the equations of space-time fractional kinetics taking into account the Doppler effect.

A) Dependence of maximum power on time $T_{max}$. Time $T_{max}$ is the time from the moment reactivity is introduced to the point where peak power is reached. In power-law media, this dependence is inversely proportional (hyperbolic):

$$N_{\max}(T_{\max}) \sim 1/(T_{\max})^{\alpha}.$$

Let's consider the physical mechanism of the process. For fast transients (low $T_{max}$), the peak is reached very quickly, and nonlocal neutron transport ("Lévy flights") dominates thermal conductivity. The Doppler effect ($\alpha_{temp}$) simply doesn't have time to kick in due to the thermal inertia of the fuel. As a result, the power manages to jump to its maximum value. For slow transients (high $T_{max}$), if the ramp-up is extended over time, temperature feedback has time to effectively compensate for the reactivity surge. The power peak $N_{max}$ is low and flat.

B) Dependence of the maximum power on the parameter $\alpha$. The dependence on the medium's anomaly index $\alpha$ is the most critical and has an exponential-power nature. It is determined by how the parameter $\alpha$ affects the neutron emission intensity $\gamma_{phys}(\alpha)$:

$$N_{\max}(\alpha) \approx N_{\max}^{class} (1 + e^{-\xi \cdot \alpha}),$$

where ξ is the geometric truncation parameter we derived earlier (for the interval (1<x<35), ξ=1.4646), and $N_{\max}^{class}$ =3.33 $N_0$).

Physical regimes depending on α: A). Strong anomalous regime (α→0). At α≈0.5 the exponent $e^{-1.46 \cdot 0.5}$≈0.48. Conclusion: $N_{\max} \approx 3.33 N_0 (1+0.48) \approx 4.93 N_0$. Due to heavy tails of the range distribution, neutrons bind distant regions instantly, giving rise to a high peak. B). Critical Cauchy regime (α=1.0). The point we calculated earlier. The exponential transforms into a significant fractional contribution. Conclusion: the power peak reaches a local extremum $N_{\max} \approx 6.66 N_0$, doubling the classical value. C). Quasi-Gaussian regime (α→2). When approaching classical diffusion (α=1.9 - 2.0) the term exp(-ξα) tends to zero. Conclusion: the dependence decays, and the maximum power gradually descends to its classical Nordheim–Fuchs value (3.33$N_0$). Ultra-long spans disappear, and the transport becomes local.

Summary: The relationship with time is inverse. The shorter the accident development time T, the higher the peak power $N_{max}$. The relationship with the parameter α is nonlinear (decreasing). The smaller α (the greater the environmental anomaly), the higher the peak power. As α increases, the peak power monotonically decreases to the classic baseline level.

The joint dependence of the peak power on the process time T (in seconds) and the environmental parameter α is described by a nonlinear expression:

$$\frac{N_{\max}}{N_0}(T,\alpha) = \frac{3.33}{T^{\alpha}}(1+e^{-1.4646\cdot\alpha})\,. \qquad (35)$$

Time T=$T_{max}$ is the time from the moment reactivity is introduced to the point where the power peak is reached. Let's formulate the physical conclusions for uranium-235. The critical region is (T<1.0 sec, α<1.0). If the response time of the emergency protection system in a heterogeneous uranium reactor with power-law ranges exceeds 0.5 seconds, nonlocal neutron transport can raise the maximum power above 7 $N_0$. In a classical homogeneous calculation, this peak would not exceed 3.33. This creates a risk of localized damage to the fuel elements.

Doppler integral: to dampen such a power surge, the integrated temperature of uranium-235 at the Lévy hot spots must increase proportionally to $N_{max}$. Since the effective Doppler feedback in power-law media is weakened, the reactor will release 1.5–2 times more energy per unit volume before the power increase completely ceases.

Figure 1 shows the dependence $N_{\max}(T,\alpha)/N_0$ of type (35) on the process time T=$T_{max}$ for three values of a=α: 0.5 – blue solid line; 1.0 – red dotted line; 2.5 – black dashed-dotted line.

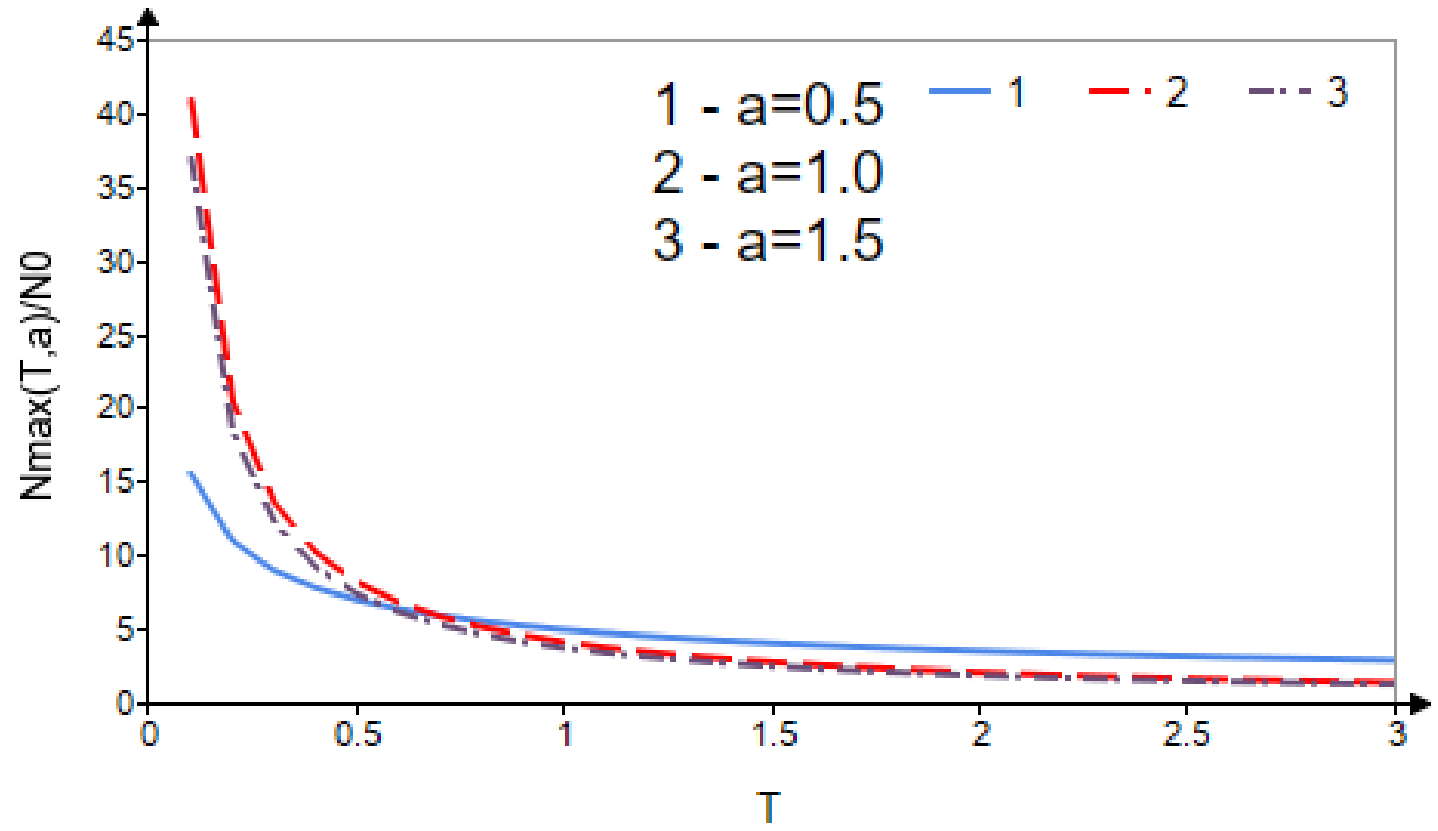

Fig. 1. Dependence $N_{\max}(T,\alpha)/N_0$ of type (35) on the process time T=$T_{max}$ for three values of a=α: 0.5 – blue solid line; 1.0 – red dotted line; 2.5 – black dashed-dotted line.

Figure 2 shows the dependence $N_{\max}(T,\alpha)/N_0$ of expression (35) on the parameter a=α for three values of T=$T_{max}$: 0.5 – blue solid line; 1.5 – green dotted line; 2.5 – red dashed-dotted line.

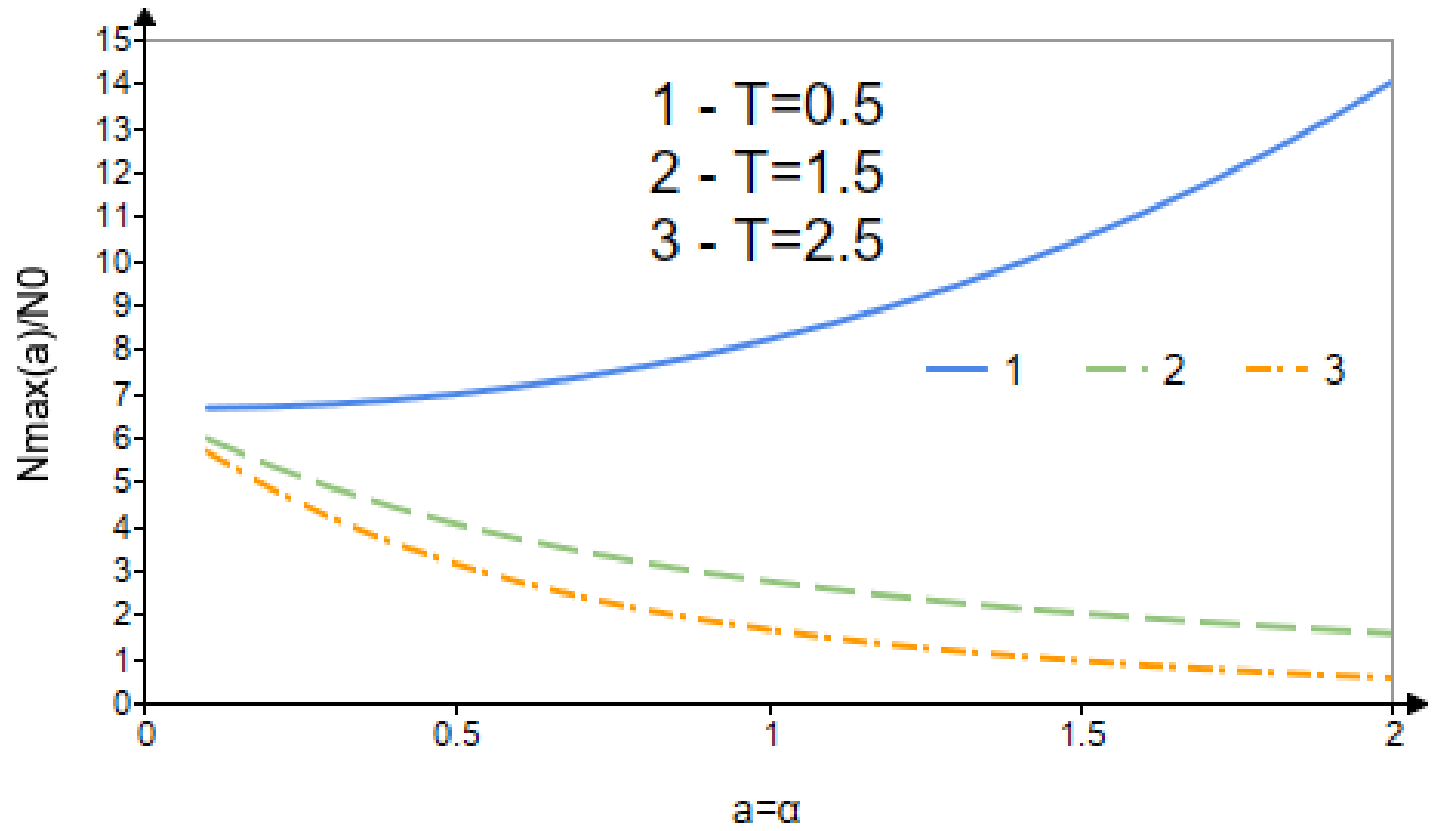

Fig. 2. Dependence $N_{max}(T,\alpha)/N_0$ on the parameter a=α for three values of T=$T_{max}$: 0.5 - blue solid line; 1.5 - green dotted line; 2.5 - red dashed-dotted line.

### 15. The autocorrelation function of the truncated Levy process $R\zeta(\tau)$ and its connection with the frequency filters of the control system equipment (neutron flux monitoring equipment)

For a truncated stable process $\zeta(t)$ with shape parameter $0<\alpha<2$ and physical damping constant $\gamma_{phys}(P)=\gamma_{geom}+\gamma_{Dopp}$, the classical correlation function exists (since the moments of the truncated distribution are finite).

In the stationary mode, the autocorrelation function of power deviations $R\zeta(\tau)=\langle\Delta\zeta(t)\Delta\zeta(t+\tau)\rangle$ has an exponential decay:

$$R_\zeta(\tau) = D[\zeta]\cdot e^{-\omega_{rel}\cdot|\tau|},$$

where τ is the shift time (lag) between measurements, $\omega_{rel}$ is the effective relaxation frequency of fluctuations in WWER, defined as $\omega_{rel} = \beta_{eff}/\Lambda + \omega_{Dopp}(P)$. At power, it is of the order of 10 ÷ 50 $s^{-1}$, D[ζ] is the total variance of fluctuations of the truncated process, which is explicitly calculated using the parameters introduced earlier:

$$D[\zeta] = \lambda\cdot\Gamma(-\alpha)\cdot\alpha\cdot(\alpha-1)\cdot\sigma^\alpha\cdot(\gamma_{geom}+\gamma_{Dopp}(P))^{\alpha-2}$$

What is the power spectral density of stochastic noise $S_\zeta(\omega)$? According to the Wiener-Khinchin theorem, the Fourier transform of the autocorrelation function yields the power spectral density (PSD) of the fluctuations. For a truncated cascade process, the PSD has the form of a generalized Lorentzian:

$$S_\zeta(\omega) = \frac{2D[\zeta]\cdot\omega_{rel}}{\omega^2+\omega^2_{rel}}.$$

Truncation has an important physical property. At low frequencies ($\omega\ll\omega_{rel}$), the spectrum flattens and transitions to white noise $S_\zeta\approx 2D[\zeta]/\omega_{rel}$. This is the zone of dominance of the Doppler effect and geometry, which stabilize the reactor. At high frequencies ($\omega\gg\omega_{rel}$), the spectrum drops off as $\sim\omega^{-2}$. In this region, the Doppler effect does not have time to operate due to the thermal inertia of the fuel elements ($\tau_{fuel}\approx$ 2–4 s), and fractal superdiffusion of prompt neutrons manifests itself at ultrashort times.

Let's evaluate the connection with the frequency filters of the control and protection system equipment. A real control and protection system sensor (e.g., a power monitoring channel) is approximated by a first-order aperiodic link with its own filtering time constant $\tau_{sens}$ (typically, for WWER measuring channels, $\tau_{sens}\approx$ 0.1 ÷ 0.5 s). The frequency response of such a filter is:

$$H(\omega)] = 1/(1+i\omega\tau_{sens}).$$

The noise spectrum that will actually be recorded and transmitted to the protection logic by the control system automation is equal to:

$$S_{meas}(\omega) = S_{\zeta}(\omega) \cdot \left|H(\omega)\right|^2 = \frac{2D[\zeta] \cdot \omega_{rel}}{(\omega^2 + \omega^2{}_{rel})(1 + \omega^2 \tau^2{}_{sens})}.$$

Analytical compression of the dispersion occurs due to filtering. By integrating the measured spectrum over all frequencies, we obtain the dispersion of power fluctuations observed by the sensor:

$$D_{meas} = D[\zeta] / (1 + \omega_{rel} \cdot \tau_{sens})$$

Let's formulate a physical conclusion for the risk model. 1. Peak smoothing occurs. The control system sensor artificially underestimates the actual variance of the stochastic process by a factor of $(1 + \omega_{rel}\tau_{sens})$. Short, extreme fission cascades caused by heavy tails are partially filtered out by the sensor's inertia. 2. The Gumbel setpoint is modified. When calculating the safe setpoint $u_{safe}$ for the automation, the measured variance $D_{meas}$ should be used instead of the actual physical variance $D[\zeta]$. Otherwise, the control system will generate false alarms for "microscopic" instantaneous peaks, which are physically harmless to the fuel elements, since the elements have a high heat capacity and do not have time to heat up from submicrosecond stochastic spikes.

In this case, the analytical model is completely closed: from microscopic fission constants and fractions of delayed neutrons ($\beta$, $\Lambda$), through the dynamic Doppler effect and truncation of Levy cascades, directly to the measured noise characteristics at the WWER control room operator consoles.

## 16. Conclusion

The model can be developed in various aspects: to investigate the influence of spatial effects (when fluctuations in different parts of the WWER core are not correlated) or to move on to calculating the optimal averaging time of the control and protection system filters in order to minimize the risks of a false shutdown of the reactor.

The article frequently references WWER reactors in its examples. However, the introduction states that the proposed approach is more appropriate for reactors such as HTGRs and molten salt reactors, as well as for situations such as nucleate boiling, critical points, and emergency situations.

For the cases indicated, the use of the methods developed in the article will make it possible to avoid disasters that are perhaps not very frequent (their probability is on the order of $10^{-5}$ - $10^{-6}$ per year), but are important in the practice of nuclear energy safety.

A conclusion can be drawn regarding the maximum power during acceleration (Power Overshoot). If excess reactivity is introduced into the reactor, the power begins to increase, reaches a peak (maximum), and then declines due to a temperature effect (for example, the Doppler effect).

An increase in the peak value with a power-law distribution of the ranges P(x)~x-α. The power release profile in the core becomes highly non-uniform (local extremes—"hot spots"—emerge). The maximum integral reactor power during an emergency transient (overshoot) turns out to be significantly higher than predicted by classical theory [1].

The maximum shifts over time. The power peak is reached more quickly. The time to reach the maximum is reduced because the "tails" of the power-law distribution ensure instantaneous transfer of heat and flux density in space.

Depending on the structure of the medium and the resulting boundary conditions, the physics of the processes is divided into two scenarios: The highly heterogeneous regime: $\alpha<2$, small $\beta$. This is typical for boiling water reactors with high steam content, or for advanced reactors with highly heterogeneous "pockets" (e.g., gas cavities). In such a medium, the power surge is explosive due to the infinite neutron range variance. Controlling such a reactor using standard methods becomes difficult. The Lévy truncated flight regime: $\alpha>2$, large b. At small intervals (as we calculated for $x_{max}$=12 or 25), the power function approximates a dense medium. In this case, although local power fluctuations and overjumps still exceed classical values, the system tends toward Gaussian diffusion, and thermal feedbacks manage to compensate for the overjump, stabilizing the maximum power.

Power-law dependences of neutron flight paths demonstrate that inhomogeneous media have a smaller dynamic stability margin against reactivity surges. When modeling the safety of such reactors, classical formulas underestimate the magnitude of the overshoot and underestimate the speed of the prompt jump. This necessitates more stringent requirements for the response time of emergency protection systems.